\documentstyle[12pt,epsfig]{article}
\begin{document}
\title{ Unitarity constraints on scalar parameters of the 
Standard and Two Higgs Doublets Model.\footnote{To be published in the 
proceedings of  ``Noncommutative Geometry, 
Superstrings and Particle Physics'' Workshop, Univ-Rabat, Morocco, 
June 16-17 2000.}}
\vspace{2mm}
\author{ A. Arhrib \\ [3mm]  
Physics Department, National Taiwan University, \\
Taipei Taiwan 10764 R.O.C }
\maketitle
\begin{abstract}
Unitarity constraints on the scalar parameters
both for the Standard Model and the general Two Higgs
Doublet Model (THDM) are examined.
In the case of the THDM with an exact discrete symmetry transformation,
we show that the mass of the lightest CP-even Higgs boson ($M_h$)
and  $\tan\beta$ are strongly correlated and consequently
a strong lower bound can be put on $M_h$ for large $\tan\beta$.
It is also shown that the inclusion of the discrete symmetry
breaking term relaxes the aforementioned correlation.
\end{abstract}

\newpage
\section{Introduction}
The Standard Model (SM) of electroweak
interactions \cite{Wein} is in complete agreement
with all precision experimental data (LEP,Tevatron, SLD).
What remains to be discovered is the standard Higgs boson ($\phi^0$), 
which is  the major goal of present and future searches at colliders.
From the $95\%$ CL upper limit obtained from
the global Standard Model fit \cite{osaka2}, 
it seems that the existence of a relatively light
Higgs boson is favored ($ M_{\phi^0} < 210$ GeV at 95\% confidence level),
which is a feature of Supersymmetric theories.
If we believe in SM, the global Standard Model
fit tell us that Higgs boson particle is going to be seen very soon. 
Direct searches at LEP show no clear evidence for the Higgs boson,
and imply a lower limit  $m_H > 112.3$ GeV at the $95\% $ CL
\cite{osaka} (Note that this is still a preliminary limit, 
the previous combined limit is  $m_H > 107.9$ GeV). 
Last September exciting new results from LEP experiments 
has postponed by one month the LEP program. 
At the end of this extra period, the data 
collected by the four experiments of LEP are 
compatible with a production of Higgs boson in 
association with Z boson but also compatible with 
other known process. The four experiment of LEP agreed 
that a further run with about 200 pb$^{-1}$ per experiment at a 
center--of--mass energy of 208.2 GeV would be enable the four experiments to
establish a 5$\sigma$ discovery \cite{www}. If there is no
extra-extension for LEP run, we have to wait until either
the RunII at Tevatron \cite{tevatronH} may aid in the interpretation of the 
signal or the LHC, which is able to discover such a light Higgs, 
start up \cite{LHC}.

Constraints on $M_{\phi^0}$
can be obtained by making
additional theoretical assumptions. These include
i) the triviality bound \cite{dashen}, ii) Vacuum stability requirement
and iii) demanding that unitarity of the S-matrix is not violated.  

The triviality bound follows from the fact that
the quartic coupling of a pure $\lambda \Phi^4$ scalar theory
increases as a function of the momentum scale, causing the 
theory to become non-perturbative at the so-called Landau pole.
The $\lambda \Phi^4$ theory will then have $\lambda(Q)=0$ 
at some scale $Q$ which renders the theory trivial.

By requiring at some large scale
$\Lambda$ (where new physics is supposed to
enter) that the quartic coupling $\lambda(\Lambda)$
satisfies the relation $\frac{1}{\lambda(\Lambda)}>0$
(which means that $\lambda(\Lambda)$ remains  non-vanishing and so the
theory is non-trivial) one
can obtain an upper bound on the Higgs boson mass.
For example, if this large scale $\Lambda$ approaches
the grand unification scale $10^{15}$ GeV one can find a strong approximate
upper bound on the Higgs boson mass \cite{quiros} $M_{\phi^0}<160$ GeV
\footnote{The above bound changes if we couple the theory to
fermions and gauge bosons.}

The  condition of vacuum stability requires that $\lambda$ 
remains positive at
a large scale in order to have spontaneous symmetry breaking;
if $\lambda$ becomes negative then the potential is unbounded from
below and has no minimum. Including the 2 loop renormalisation
group improved effective potential, the vacuum stability requirement
gives (if the Standard model is valid up to scale $10^{16}$ GeV)
\cite{scher, quiros}:
$M_{\phi^0} > 130.5 + 2.1 (M_t - 174)$ GeV.
Finally, unitarity arguments lead
to an upper bound on the Higgs boson mass \cite{uni}, and this will
be covered in more detail in section 2.

In recent years there has been growing
interest in the study of extended
Higgs sectors with more than one doublet \cite{Gun}.
The simplest extension of the MSM is the Two Higgs
Doublet Model (THDM), which is formed by adding an extra complex
$SU(2)_L\otimes U(1)_Y$ scalar doublet to the MSM Lagrangian.
Motivations for such a structure include CP--violation in the Higgs
sector, supersymmetry, and a possible solution to the cosmological
domain wall problem \cite{preskill}. In particular, the Minimal
Supersymmetric Standard Model (MSSM)
\cite{Gun} takes the form of a constrained THDM.
The two most popular  versions of THDM 
are classified as type I and type II, differing in how
the Higgs bosons are coupled to the fermions although
both versions possess identical particle spectra
after electroweak symmetry  breaking. From the eight degrees
of freedom initially present in the two Higgs doublets, three
correspond to masses of the longitudinal gauge bosons, leaving
five degrees of freedom which manifest themselves as five
physical Higgs particles: Two charged Higgs $H^\pm$, two
CP-even $H^0$, $h^0$ and one CP-odd $A^0$. Until now no Higgs boson has
been discovered, and from the null searches one can derive direct
and indirect bounds on their masses. The latest such limits are:
for the charged Higgs boson the LEP combined result is
$m_{H^\pm}>77.5$ GeV \cite{osaka}, and for the
neutral Higgs OPAL collaboration \cite{opal1} has made the
following scan of the parameter space as:
$1<m_h<100$, $5<m_A<2$ TeV, $-\pi/2 < \alpha <0$ and $0.4 <\tan\beta <58$
finding that the region ( $1< m_h <44$ GeV and
$12 < m_A < 56$ GeV ) is excluded at 95\% independent of
$\alpha$ and $\tan\beta$.

We note here that tree-level unitarity constraints for
the THDM scalar potential constrained by the exact discrete
symmetry $\Phi_i \to -\Phi_i$  were studied in
\cite{malampi, kanemura}. When deriving constraints from unitarity,
\cite{malampi} considered only seven elastic scattering processes
$S_1S_2\to S_1 S_2$ (where $S_i$ is a Higgs scalar)
while \cite{kanemura} considered
a larger scattering ($S$) matrix. In \cite{kanemura}, upper bounds
on the Higgs masses were derived, in particular $M_h\le 410$ GeV
for $\tan\beta=1$, with the bound becoming
stronger as $\tan\beta$ increases. 
In section. 3, at the same 
footing as we will present for the MSM in section. 2,  
we improve the aforementioned THDM studies by including
the full scalar $S$ matrix which includes channels which were 
absent in \cite{kanemura}. In addition, we also show graphically 
the strong correlation between $M_h$ and $\tan\beta$ both in the 
case where the discrete symmetry $\Phi_i \to -\Phi_i$ is exact 
and also in the case where it is broken by dimension 2 terms: 
$\lambda_5 \Re (\Phi_1^+\Phi_2)$.

The paper is organized as follows. In Section 2 we present the unitarity
approach we will be using and review the unitarity constraints
in the framework of the MSM. Section 3 contains a short review of
the THDM potential and the analytical unitarity constraints.
In Section 4 we present our numerical results for the
cases of $\lambda_5=0$ and $\ne 0$, while Section
5 contains our conclusions. The appendix contains pure scalar
quartic interactions both for MSM and for THDM.

\renewcommand{\theequation}{2.\arabic{equation}}
\setcounter{equation}{0}
\section*{2. Unitarity  approach}
We will study the unitarity constraints by computing
the scalar scattering processes, $S_1 S_2 \to S_3 S_4$, i.e. both
elastic and inelastic channels.
In terms of the partial wave decomposition, the amplitude ${\cal M}$
of a scattering $S_1 S_2 \to S_3 S_4$ can be written as:
\begin{eqnarray}
{\cal M}(s,t,u) = 16 \pi \sum_{l=0}^{\infty} (2 l +1)
P_l(\cos\theta) a_l(s)  \label{partial}
\end{eqnarray}
Where $s$, $t$, $u$ are the Mandelstam variables,
$a_l(s)$ is the spin $l$ partial wave
and $P_l$ are Legendre Polynomials.
The differential cross section for $S_1 S_2 \to S_3 S_4$ is given by:
$$\frac{d\sigma}{d\Omega}=\frac{1}{64 \pi^2 s} |{\cal M}|^2$$
Using the fact the the Legendre polynomials are orthogonal, the cross
section becomes:
$$ \sigma= \frac{16 \pi}{s}\Sigma_{l=0}^{\infty} (2l +1) |a_l|^2 $$
The optical theorem together with expression for the total cross
section given above leads to the following unitarity constraint:
$$ \Re(a_l)^2 + \Im (a_l)^2 = |a_l|^2 = \Im (a_l) \qquad
\mbox{for all} \ \ \  \ l  $$
The above equation is nothing more than an equation for a circle 
in the plane 
$(\Re(a_l), \Im (a_l) )$ with radius $\frac{1}{2}$ and center $(0,\frac{1}{2})$.
It can be shown easily from the graphical representation
of this circle that:
$$|\Re(a_l)| < \frac{1}{2} \qquad \mbox{for all } \  \ \ \ l $$

The partial wave $a_l(s)$ can be inversed from eq.\ref{partial} and one 
finds:
$$a_l(s) = \frac{1}{32 \pi} \int_{-1}^{1} d(\cos\theta)
P_l(\cos \theta) {\cal M} (s,t,u) $$
If we limit ourselves to the $J=0$ s--wave amplitude $a_0(s)$,
for vanishing external masses of $S_{1,2,3,4}$,
$a_0(s)$ take the following form:
\begin{eqnarray}
a_0(s) & = & \frac{1}{16 \pi} [ Q + \{ T_h^{12} 
T_h^{34}\frac{1}{s-M_h^2}
 - \frac{1}{s} (c_t T_h^{13}T_h^{24}+c_u T_h^{14}T_h^{23})
\mbox{ln}(1+\frac{s}{m_h^2})\}]\label{a0}
\end{eqnarray}
Where $Q$ is the four point vertex for $S_1 S_2\to S_3 S_4$ and
$T_{h}^{ij}$ are the trilinear vertex $hS_iS_j$,
$c_t=1\ \mbox{or}\ 0 $ (resp
$c_u=1\ \mbox{or}\ 0)$
for processes with or without t-channel (resp for process
with or without u-channel).
\\ The first term in the bracket in the r.h.s of eq.\ref{a0}
is the contribution of the quartic coupling $Q=S_1S_2S_3S_4$ to
the amplitude ${\cal M}$; the second term is the contribution
of the s-channel diagram with exchange of particle $h$
and the third term is the contribution of t and u channels.

In very high energy collisions, it can be shown from eq.\ref{a0} that
the dominant contribution to the amplitude of the two-body scattering
$S_1 S_2 \to S_3 S_4$ is the one which is mediated by the quartic
coupling. Those contributions mediated by trilinear couplings
are suppressed on dimensional grounds.
Therefore the unitarity constraint $|a_0|\leq 1/2$ reduces to
the following constraint on the quartic coupling,
$|Q( S_1 S_2  S_3 S_4)|\leq 8 \pi$.
In what follows our attention will be focused on the 
quartic couplings.

To constrain the scalar potential parameters one can demand
that tree-level unitarity is preserved in a variety of scattering
processes. This corresponds to the requirement
that the $J=0$ partial waves ($a_0$) for
{\underline scalar-scalar}, {\underline gauge boson-gauge boson}
and {\underline gauge boson-scalar} scattering
satisfy $|a_0|<1/2$ in the high-energy limit.
At very high-energy,
the equivalence theorem \cite{equivalence} states that
the amplitude of a scattering process involving longitudinal gauge
bosons $V_{\mu}^{\pm,0}$
may be approximated by the scalar amplitude in which gauge bosons are
replaced by their corresponding Goldstone bosons $G^{\pm,0}$.
We conclude that unitarity constraints can be implemented
by solely considering pure scalar scattering.

Let us apply all these features to the non-coupled process
$W_L^+W_L^- \to W_L^+W_L^-$ in the MSM, where the subscript $L$
denotes the longitudinal polarisation states.
As stated above, at high energy the amplitude of
$W_L^+W_L^- \to W_L^+W_L^-$ is approximated by $G^+G^- \to G^+G^-$ 
(where
$G^\pm$ is the charged goldstone associated with $W^\pm$) whose
dominate contribution comes the quartic coupling,
$G^+G^-G^+G^-= M_{\phi^0}^2/v^2$, where $v$ is fixed by the 
electroweak scale as $v^2=(2\sqrt{2} G_F)^{-1}$. The unitarity constraint
$|Q|\leq 8 \pi$ gives the following upper bound on the Higgs boson's mass:
$$ M_{\phi^0}^2 < 8\pi v^2=M_{LQT}^2=
\frac{4\pi}{\sqrt{2} G_F}=(870)^2 \mbox{GeV}^2 $$
Where $M_{LQT}$  is the bound deduced by Lee, Quigg and
Thacker \cite{uni}.\\
In fact the channel $W_L^+W_L^-$ considered above is coupled with
the following channels: $Z_LZ_L/\sqrt{2}$, $\phi^0\phi^0/\sqrt{2}$
and $Z_L\phi^0$ (the factor $\sqrt{2}$ accounts for identical
particle statistics). Together with those four neutral initial states
we consider also the 2 charged channels $W_L^+\phi^0$ and
$W_L^+Z$. Note that due to charge conservation the two
charged channels are not coupled to the four charged channels.
Taking into account all the above channels, the scattering amplitude is
given by $6\times 6$ matrix which is diagonal by block: $4\times 4$ block
for the 4 neutral channels and $2\times 2$ block for 2 charged channels.
At high energies, the matrix elements are dominated by the
quartic couplings and so the full matrix
in the basis ($W_L^+W_L^-$, $Z_LZ_L/\sqrt{2}$, $\phi^0\phi^0/\sqrt{2}$,
$Z_L\phi^0$,$W_L^+\phi^0$,$W_L^+Z$) takes the following form:
\begin{eqnarray}
a_0= \frac{M_{\phi^0}^2}{v^2} \left(
\begin{array}{cccccc}
1 & \frac{\sqrt{2}}{4} & \frac{\sqrt{2}}{4} & 0 & 0 & 0 \\
 \frac{\sqrt{2}}{4} & \frac{3}{4} & \frac{1}{4} & 0 & 0 & 0 \\
 \frac{\sqrt{2}}{4} & \frac{1}{4} & \frac{3}{4} & 0 & 0 & 0 \\
 0 & 0 & 0 & \frac{1}{2} & 0 & 0 \\
 0 & 0 & 0 & 0 & \frac{1}{2} & 0 \\
 0 & 0 & 0 & 0 & 0 & \frac{1}{2}\end{array}  \right) \label{msm}
\end{eqnarray}
This matrix can be read easily from the appendix A.1 in which we have
listed the pure scalar quartic interactions in the MSM.\\
The requirement that the largest
eigenvalues\footnote{In the case of this
matrix the largest eigenvalues is 3/2} of $a_0$ respects the
unitarity constraint yields\cite{marciano}:
$$ M_{\phi^0}^2 < \frac{2}{3} 8\pi v^2=
\frac{2}{3} M_{LQT}^2 \approx (710)^2 \mbox{GeV}^2 $$
We conclude that the inclusion of the complete
set of scattering channels (charged and neutral ones)
into the analysis, leads more or less to a stronger unitarity 
constraint on the Higgs boson mass $M_{\phi^0} < 710 \mbox{GeV}$.
The unitarity requirement tells us that if the Higgs boson mass is
above the unitarity constraint then the standard model becomes
non-perturbative and some new physics should appear at some high
scale to restore the unitarity of the theory.

\renewcommand{\theequation}{3.\arabic{equation}}
\setcounter{equation}{0}
\section*{3. THDM scalar potential and Unitarity constraints}
\subsection*{3.1 THDM scalar potential}
The most general THDM scalar potential which is renormalizable,
gauge invariant and CP invariant depends on ten parameters,
but such a potential can still break CP spontaneously \cite{Lee}.
In order to ensure that tree-level flavor changing neutral currents
are eliminated, a discrete symmetry ($\Phi_i\to -\Phi_i$, where
$\Phi_i$ is a scalar doublet) may be imposed on the lagrangian
\cite{sym}, which reduces the number of free parameters to 6.
The resulting potential was considered in \cite{sher}, and is
referred to as $V_A$ in \cite{brucher}.
We shall be concerned with the potential described in \cite{Gun}
which is equivalent to $V_A$ plus a term which
breaks the discrete symmetry
(parameterized by $\lambda_5$) and contains 7 free parameters.
Such a potential does not break CP spontaneously or explicitly
\cite{sher},\cite{branco} provided that all the parameters are real.

It has been shown
\cite{Geo} that the most general THDM scalar potential which is
invariant under $SU(2)_L\otimes U(1)_Y$ and conserves CP is given by:
\begin{eqnarray}
 V(\Phi_{1}, \Phi_{2})& & =  \lambda_{1} ( |\Phi_{1}|^2-v_{1}^2)^2
+\lambda_{2} (|\Phi_{2}|^2-v_{2}^2)^2+
\lambda_{3}((|\Phi_{1}|^2-v_{1}^2)+(|\Phi_{2}|^2-v_{2}^2))^2
+\nonumber\\ [0.2cm]
&  & \lambda_{4}(|\Phi_{1}|^2 |\Phi_{2}|^2 - |\Phi_{1}^+\Phi_{2}|^2  )+
\lambda_{5} (\Re(\Phi^+_{1}\Phi_{2})
-v_{1}v_{2})^2+ \lambda_{6} [\Im(\Phi^+_{1}\Phi_{2})]^2
\label{higgspot}
\end{eqnarray}
where $\Phi_1$ and $\Phi_2$ have weak hypercharge Y=1, $v_1$ and
$v_2$ are respectively the vacuum
expectation values of $\Phi_1$ and $\Phi_2$ and the $\lambda_i$'s
are real--valued parameters.

$$\Phi_i=\left( \begin{array}{c}
w_i^+\\
v_i+\frac{h_i + iz_i}{\sqrt{2}}
\end{array}\right) $$
This potential violates the discrete symmetry
$\Phi_i\to -\Phi_i$ softly by the dimension 2 term
$\lambda_5 \Re(\Phi^+_{1}\Phi_{2})$ and has the same
general structure of the scalar potential of the MSSM.
One can prove easily that for $\lambda_5=0$ 
the exact symmetry $\Phi_i \to -\Phi_i$ is recovered.

After electroweak symmetry breaking, the W and Z gauge
bosons acquire masses 
given by  $m_W^2=\frac{1}{2}g^2 v^2$ and
$m_Z^2= \frac{1}{2}(g^2 +g'^2) v^2$,
where $g$ and $g'$ are the $SU(2)_{weak}$ and
$U(1)_Y$ gauge couplings and
$ v^2= v_1^2 + v_2^2$. The combination $v_1^2 + v_2^2$
is thus fixed by the electroweak
scale through $v_1^2 + v_2^2=(2\sqrt{2} G_F)^{-1}$,
and we are left with 7 free parameters in eq.(\ref{higgspot}),
namely the $(\lambda_i)_{i=1,\ldots,6}$'s and
$\tan\beta=v_2/v_1$. Meanwhile,  three of the eight degrees
of freedom  of the two Higgs doublets correspond to
the 3 Goldstone bosons ($G^\pm$, $G^0$) and 
the remaining five become physical Higgs bosons:
$H^0$, $h^0$ (CP--even), $A^0$ (CP--odd)
and $H^\pm$. Their masses are obtained as usual
by the shift $\Phi_i\to \Phi_i + v_i$ \cite{Gun} and read     
\begin{eqnarray}
&& m_{A^0}^2=\lambda_6 v^2\  ;  \  m_{H^{\pm}}^2=
\lambda_4 v^2\  \mbox{and} \
 m_{H^0,h^0}^2=\frac{1}{2} [ A+C \pm \sqrt{(A-C)^2+4B^2} ] \label{higgsmass}
\end{eqnarray}
where
\begin{eqnarray}
&& A=4 v_1^2 (\lambda_1+\lambda_3)+v_2^2\lambda_5\ \ , \ \  B= v_1 v_2
(4\lambda_3+\lambda_5)\ \ \mbox{and} \ \
C=4 v_2^2 (\lambda_2+\lambda_3)+v_1^2\lambda_5 \nonumber
\end{eqnarray}
The angle $\beta$ given by $\tan \beta= v_2/v_1$ defines 
the mixing leading to
the physical $H^\pm$ and $A^0$ states, while the mixing angle $\alpha$ 
associated to $H^0, h^0$ physical states is given by 
(See  \cite{Gun} for details.)
\begin{eqnarray}
&& \sin 2 \alpha=\frac{2 B}{\sqrt{(A-C)^2+4 B^2} }\ , \ \ \
\cos 2 \alpha=\frac{A-C}{\sqrt{(A-C)^2+4 B^2} }
\label{alph}
\end{eqnarray}
It will be more suitable for the forthcoming discussion to trade the five
parameters $\lambda_{1, 2,4,5,6}$ for the 4 Higgs masses 
and the mixing angle $\alpha$. From now on we will take 
the physical Higgs masses,
$m_{H^0}, m_{h^0}, m_{A^0}, m_{H^\pm} $, the mixing angles $\alpha, \beta$ and
the coupling $\lambda_5$ as the 7 free parameters. 

It is then straightforward algebra to invert equations (\ref{higgsmass})
through (\ref{alph}), and get the $\lambda_i$'s in terms of
this new set of parameters.

\begin{eqnarray}
& & \lambda_4=\frac{g^2}{2 m^2_W} m_{H^\pm}^2 \ \ ,  \ \
\lambda_6=\frac{g^2}{2 m^2_W} m_{A}^2  \label{lambda46}
\ \ , \ \  \lambda_3=\frac{g^2}{8 m^2_W}
\frac{\mbox{s}_\alpha \mbox{c}_\alpha}{ \mbox{s}_\beta \mbox{c}_\beta }
(m_H^2-m_h^2)\ -
\  \frac{\lambda_5}{4} \label{lambda5}  \\
& & \lambda_1 = \frac{g^2}{8 \mbox{c}_\beta^2 m^2_W}
[ \mbox{c}_\alpha^2 m^2_H+
\mbox{s}_\alpha^2 m^2_h -
\frac{\mbox{s}_\alpha \mbox{c}_\alpha}{\tan\beta}(m^2_H - m^2_h)]
 -\frac{\lambda_5}{4}(-1 + \tan^2\beta) \label{lambda1} \\ & &
\lambda_2 = \frac{g^2}{8 \mbox{s}_\beta^2 m^2_W} [ \mbox{s}_\alpha^2 
m^2_H+
\mbox{c}_\alpha^2 m^2_h -
\mbox{s}_\alpha \mbox{c}_\alpha \tan\beta(m^2_H - m^2_h)]
 -\frac{\lambda_5}{4}(-1 + \frac{1}{\tan^2\beta} ) \label{lambda1}
\end{eqnarray}

\subsection*{3.2 Unitarity constraints}
To constrain the scalar potential parameters of the
THDM one can demand that tree-level unitarity is preserved
in a variety of scattering processes: scalar-scalar scattering,
gauge boson-gauge boson scattering and scalar-gauge boson scattering.
We will follow exactly the technique developed in section.2
and therefore we limit ourselves to pure scalar scattering
processes dominated by quartic interactions.

In order to derive the unitarity constraints on the scalar masses
we will adopt the technique introduced in \cite{kanemura}.
It has been shown in previous works \cite{our}
that the quartic scalar vertices written
in terms of physical fields $H^\pm$, $G^\pm$, $h^0$, $H^0$, $A^0$
and $G^0$, are very complicated functions of $\lambda_i$,
$\alpha$ and $\beta$. However the quartic vertices
(computed before electroweak symmetry breaking)
written in terms of the non-physical fields $w_i^\pm$ ,
$h_i$ and $z_i$ (i=1,2) are considerably simpler expressions.
The crucial point of \cite{kanemura} is the fact that
the $S$ matrix expressed in terms of the physical fields
(i.e. the mass eigenstate fields) can be transformed into
an $S$ matrix for the non-physical fields
$\varphi_i^\pm$ , $h_i$ and $z_i$
by making a unitarity transformation.
The latter is relatively easy to compute from  eq. \ref{higgspot}.
Therefore the full set of scalar scattering processes can be expressed 
as
an $S$ matrix $ 22\times 22$ composed of 4 submatrices
[ ${\cal M}_1(6 \times 6)$, ${\cal M}_2(6\times 6)$,
 ${\cal M}_3(6\times 6)$ and ${\cal M}_4(8\times 8)$] which
do not couple with each other due to charge conservation and
CP-invariance. The entries are the quartic couplings which mediate
the scattering processes.
The pure scalar quartic
interactions expressed in terms of the non-physical fields $h_i^\pm$,
$h_i$ and $z_i$ are  listed in the appendix A.2.

The first submatrix ${\cal M}_1$  corresponds to scattering whose
initial and final states are one of the following:
$(w_1^+w_2^-$,$w_2^+w_1^-$, $h_1 z_2$, $h_2z_1 
$, $z_1 z_2$, $h_1h_2)$. With the help of appendix A.2, 
one can find that ${\cal M}_1$ takes the following form:

\begin{eqnarray}
{\cal M}_1=\left(
\begin{array}{cccccc}
 2\lambda_3 +\lambda_{56}^+/2 &\lambda_{56}^- &
 -i\lambda_{64}^-/2 &i \lambda_{64}^-/2 &
 \lambda_{54}^-/2  &  \lambda_{54}^-/2  \\
\lambda_{56}^- & 2\lambda_3 +\lambda_{56}^+/2 &
i\lambda_{64}^-/2 &-i\lambda_{64}^-/2 &
\lambda_{54}^-/2  &  \lambda_{54}^-/2 \\
 i\lambda_{64}^-/2  &-i\lambda_{64}^-/2  &
2\lambda_3+\lambda_6 & \lambda_{56}^-/2 &
 0  &  0 \\
 -i\lambda_{64}^-/2  &i\lambda_{64}^-/2  &
 \lambda_{56}^-/2 &2\lambda_3+\lambda_6 &
 0  &  0 \\
\lambda_{54}^-/2  &  \lambda_{54}^-/2  & 0 &0
& 2\lambda_3+\lambda_5 & \lambda_{56}^-/2 \\
\lambda_{54}^-/2  &  \lambda_{54}^-/2 & 0 & 0 &
\lambda_{56}^-/2 & 2\lambda_3+\lambda_5
\end{array}\right)\nonumber
\end{eqnarray}
$\ $
\\
where $\lambda_{ij}^{\pm}=\lambda_i\pm \lambda_j$. With the 
help of Mathematica
we find that ${\cal M}_1$ has the following
5 distinct eigenvalues:
\begin{eqnarray}
& & e_1=2 \lambda_3 - \lambda_4 - \frac{\lambda_5}{2} + \frac{5}{2} 
\lambda_6 \nonumber\\
& & e_2=2 \lambda_3 + \lambda_4 - \frac{\lambda_5}{2} + 
\frac{1}{2}\lambda_6 \nonumber\\
& & f_+=2 \lambda_3 - \lambda_4 +  \frac{5}{2}\lambda_5 - \frac{1}{2} 
\lambda_6\nonumber\\
& & f_-=2 \lambda_3 + \lambda_4 + \frac{1}{2}\lambda_5 - 
\frac{1}{2}\lambda_6\nonumber\\
& & f_1=f_2=2 \lambda_3 + \frac{1}{2} \lambda_5 + \frac{1}{2} 
\lambda_6
\end{eqnarray}

The second submatrix ${\cal M}_2$ corresponds to scattering with 
initial and final
states one of the following: 
$(w_1^+w_1^-$, $w_2^+w_2^-$,
$\frac{z_1 z_1}{\sqrt{2}}$,
 $\frac{z_2z_2}{\sqrt{2}}$, $\frac{h_1 h_1}{\sqrt{2}}$,
$\frac{h_2h_2}{\sqrt{2}})$, where the $\sqrt{2}$ accounts for
identical particle statistics. Again, with the help of appendix A.2,
one find that ${\cal M}_2$ is given by:

\begin{eqnarray}
{\cal M}_2=\left(
\begin{array}{cccccc}
4\lambda_{13}^+ & 2\lambda_3 +\lambda_{56}^+/2 &
\sqrt{2} \lambda_{13}^+ &
{\tilde{\lambda}}_{34}/\sqrt{2} &
\sqrt{2}\lambda_{13}^+  &  {\tilde{\lambda}}_{34}/\sqrt{2}
\\
2\lambda_3 +\frac{\lambda_{56}^+}{2} &  4\lambda_{23}^+
& {\tilde{\lambda}}_{34}/\sqrt{2} &
\sqrt{2}\lambda_{23}^+ & {\tilde{\lambda}}_{34}/\sqrt{2} &
\sqrt{2}\lambda_{23}^+
\\
\sqrt{2}\lambda_{13}^+  & {\tilde{\lambda}}_{34}/\sqrt{2}
&   3\lambda_{13}^+ & {\tilde{\lambda}}_{35}/2&
\lambda_{13}^+  &   {\tilde{\lambda}}_{36}/2
\\
{\tilde{\lambda}}_{34}/\sqrt{2}  &
 \sqrt{2} \lambda_{23}^+  &   {\tilde{\lambda}}_{35}/2&
3 \lambda_{23}^+ &
 {\tilde{\lambda}}_{36}/2  &   \lambda_{23}^+
\\ \sqrt{2} \lambda_{13}^+ &
{\tilde{\lambda}}_{34}/\sqrt{2} &  \lambda_{13}^+
&{\tilde{\lambda}}_{36}/2 &
 3 \lambda_{13}^+  &  {\tilde{\lambda}}_{35}/2
\\
{\tilde{\lambda}}_{34}/\sqrt{2}  & \sqrt{2}\lambda_{23}^+
& {\tilde{\lambda}}_{36}/2 &  \lambda_{23}^+
 & {\tilde{\lambda}}_{35}/2 & 3 \lambda_{23}^+
\end{array}\right)
\end{eqnarray}
$\ $ \\
where $\tilde{\lambda}_{3i}=2\lambda_3+\lambda_i$.
The matrix ${\cal M}_2$ possesses the following 6 distinct eigenvalues:
\begin{eqnarray}
& & a_{\pm} =3 (\lambda_1 + \lambda_2 + 2 \lambda_3) \pm
\sqrt{9 (\lambda_1 - \lambda_2)^2 +
      (4 \lambda_3 +  \lambda_4 + \frac{1}{2}(\lambda_5 + 
\lambda_6))^2}\\
& & b_{\pm}=\lambda_1 + \lambda_2 +
2 \lambda_3 \pm \sqrt{ (\lambda_1 - \lambda_2)^2 +
\frac{1}{4}(-2 \lambda_4 + \lambda_5 + \lambda_6)^2}\\
& & c_{\pm}=\lambda_1 + \lambda_2 +
2 \lambda_3 \pm \sqrt{(\lambda_1 -
\lambda_2)^2 + \frac{1}{4}(\lambda_5 - \lambda_6)^2}
\end{eqnarray}

The third submatrix ${\cal M}_3$ corresponds to the basis:
$(h_1 z_1 , h_2 z_2)$ and is given by:

\begin{eqnarray}
{\cal M}_3=
\left(
\begin{array}{cccccc}
2 \lambda_{13}^+ & \frac{1}{2}\lambda_{56}^- \\
 \frac{1}{2}\lambda_{56}^-  & 2 \lambda_{23}^+
\end{array}\right)
\end{eqnarray}
$\ $ \\
The  matrix ${\cal M}_3$
possesses the eigenvalues $d_{\pm}$ and $c_{\pm}$, with
$d_{\pm}=c_{\pm}$.
All the above eigenvalues agree with those found in \cite{kanemura},
up to a factor of $1/16\pi$ which we have factorised out.
In our analysis we also include the two body scattering
between the 8 charged states: $h_1 w_1^+ $,  $h_2 w_1^+ $,
$z_1 w_1^+ $, $z_2 w_1^+$, $h_1w_2^+$, $h_2 w_2^+$, $z_1 w_2^+$, 
$z_2 w_2^+$.
The 8$\times$8 submatrix ${\cal M}_4$ obtained from the above
scattering processes is given by:

\begin{eqnarray}
{\cal M}_4=
\left(
\begin{array}{cccccccc}
 2 \lambda_{13}^+ & 0 & 0 & 0 & 0 & 
\lambda_{54}^-/2 & 0 &
-i \lambda_{64}^-/2
\\
0 & 2\lambda_3 + \lambda_4 & 0 & 0  &  \frac{1}{2}\lambda_{54}^- & 0 
&
i  \frac{1}{2}\lambda_{64}^- & 0
\\
0 & 0 & 2\lambda_{13}^+ & 0 & 0  &
 \frac{i}{2}\lambda_{64}^- & 0 &
 \lambda_{54}^-/2
\\
0 & 0 & 0 & 2\lambda_3 + \lambda_4 & -  \frac{i}{2}\lambda_{64}^- & 
0 &
 \frac{1}{2}\lambda_{54}^- & 0
\\
0 & \frac{1}{2} \lambda_{54}^- & 0 &  
 \frac{i}{2}\lambda_{64}^-
&2 \lambda_3 + \lambda_4 &
 0 & 0 &0
\\
 \frac{1}{2}\lambda_{54}^- & 0 &
- \frac{i}{2}\lambda_{64}^- & 0 & 0 & 2 \lambda_{23}^+ &
 0 & 0
\\
0 & - \frac{i}{2}\lambda_{64}^- & 0 &
 \frac{1}{2}\lambda_{54}^- & 0 & 0 & 2 \lambda_3 + \lambda_4 &  0
\\
 \frac{i}{2}\lambda_{64}^- & 0 &
 \frac{1}{2}\lambda_{54}^- & 0 & 0 &  0 & 0 & 2 \lambda_{23}^+
\end{array}\right)\nonumber
\end{eqnarray}
$\ $\\
As one can see, this matrix contains many vanishing elements, 
and the 8 eigenvalues are 
straightforward to obtain analytically. They read as follows:
$f_-$, $e_2$ , $f_{1}$, $c_{\pm}$, $b_{\pm}$ and $p_1$,
where
\begin{eqnarray}
p_1 =2 (\lambda_3 + \lambda_4 ) - \frac{1}{2}\lambda_5 -
\frac{1}{2}\lambda_6 \label{2.9}
\end{eqnarray}
As one can see, these additional channels lead only to one extra
eigenvalue, $p_1$, although we shall see that this eigenvalue plays
an important role in constraining $M_{H^\pm}$ and $M_A$.

\renewcommand{\theequation}{4.\arabic{equation}}
\setcounter{equation}{0}
\section*{4. Numerical results and discussion}
In this section we present our results for the unitarity constraints
on the Higgs masses in the THDM. All the eigenvalues are
constrained as follows:
\begin{eqnarray}
|a_{\pm}| \ , \ |b_{\pm}| \ , \ |c_{\pm}| \ , \ |d_{\pm}| \ , \ 
|f_{\pm}| \ ,
\ |e_{1,2}| \ , \ |f_{1,2}| \ , \ |p_1| \ \leq 8\pi
\label{constraint}
\end{eqnarray}
In our numerical illustrations, we will consider the 
special case of $\lambda_5=0$ and also the general 
case of  $\lambda_5\ne 0$.
\subsection*{Case of $\lambda_5=0$}
For $\lambda_5=0$ our potential is identical to those considered
in \cite{malampi,kanemura}. We improve those analyses on two accounts:

\begin{itemize}
\item [{(i)}] We have considered extra scattering channels which
leads to one more eigenvalue constraint, $p_1$, as explained in Section. 3.

\item [{(ii)}] When finding the allowed parameter space of Higgs masses
we simultaneously impose all the eigenvalue constraints.
In \cite{kanemura} only the condition $|a_+|\le 8\pi$ was applied when
deriving mass bounds.
\end{itemize}

In order to obtain the upper bounds on the Higgs masses
allowed by the unitarity constraints we vary all the Higgs
masses and mixing angles randomly over a very large parameter space.
As is reported in \cite{AAN},
we confirm the result of \cite{kanemura} 
which states that $a_+$ is
comfortably the strongest individual eigenvalue constraint. 
However, the other eigenvalues impose important 
constraints on $M_A$ and $M_{H^\pm}$. If only $|a_+|\le 8\pi$ 
is imposed we can reproduce the upper bounds on the Higgs masses given
in \cite{kanemura}, in particular their main result of $M_h\le 410$ GeV.
When all eigenvalue bounds are applied simultaneously we find improved
bounds on the Higgs masses, particularly for $M_A$ and $M_{H^\pm}$.
We note that the new eigenvalue constraint $p_1\le 8\pi$ (eq.\ref{2.9})
plays a crucial role in determining the upper bound on $M_{H^\pm}$.
For $\lambda_5=0$, we find the following upper bounds:
$$
M_{H^\pm} < 691\quad , \quad M_A < 695 \quad , \quad  M_h < 435 
\quad , \quad M_H <638 \quad (GeV) $$

Note that the bounds given above are obtained for
relatively small $\tan\beta$ (say $\tan\beta\approx 0.5$). For
large $\tan\beta$ the bound is stronger, although for the case of
$A^0$, $H^0$ and $H^{\pm}$ the $\tan\beta$ dependence is
rather gentle. Of particular interest is
the $\tan\beta$ dependence of the bound on $M_h$ which will covered
below.\\
We have to stress here that our results in the case where the discrete
symmetry is exact agree with the results obtained using 
triviality approach \cite{trivial}. 
\begin{figure}[hbt]
\centerline{\protect\hbox{\psfig{file=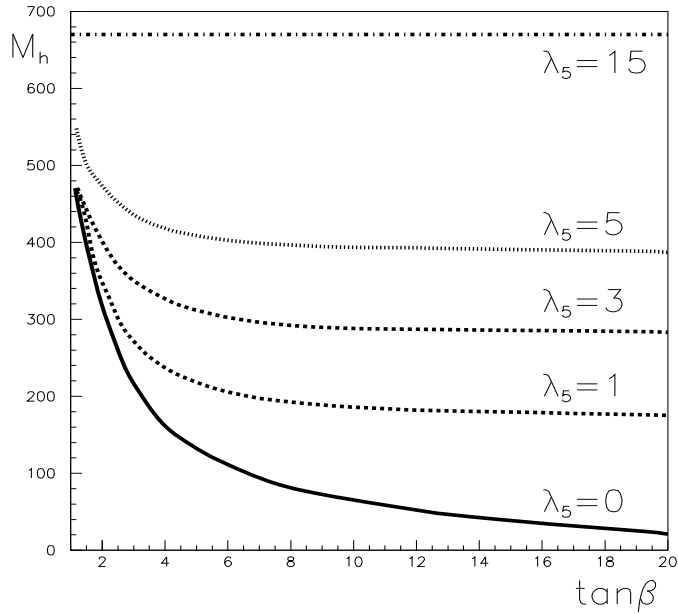,height=10cm,width=
10cm}}}
\caption{Maximum $M_h$ in GeV as a function of $\tan\beta$ for various
values of $\lambda_5$.}
\end{figure}
\subsection*{General case of $\lambda_5\ne 0$}
We now consider $\lambda_5\ne 0$ which corresponds to the
inclusion of the term which softly breaks the discrete symmetry.
Such a term was neglected in the analyses of \cite{malampi, kanemura},
and from perturbative constraints may take values $|\lambda_5|\le 8\pi$
\cite{okada}. In the graphs which follow we do not impose the
perturbative requirement $|\lambda_i|\le 8\pi$ for the remaining
$\lambda_i$.  Imposing this condition only leads to minor changes in the
numerical results which will be commented on when necessary.

We plot in Fig.1 the maximum value of
$M_h$ against $\tan\beta$ for increasing values of $\lambda_5$, imposing
all the eigenvalue constraints simultaneously as done in Section 3.1.
For the case of $\lambda_5=0$ one finds a strong correlation, with
larger $\tan\beta$ requiring smaller $M_h$.
For example, $\tan\beta\ge 7$ corresponds to $M_h\le 100$ GeV, which is
the mass range already being probed by LEPII.
However, $h^0$ in the THDM with $M_h\le 100$ GeV is not guaranteed to
be found at LEPII due to the suppression factor of 
$\sin^2(\beta-\alpha)$
for the main production process $e^+e^-\to h^0Z$.
For the case of $\lambda_5=0$ we find that values of
$\tan\beta\ge 20$ are strongly disfavoured
since they easily violate one of the unitarity constraints.
If $\lambda_5\ne 0$, Fig.1 shows that for a given $\tan\beta$,
the action of increasing $\lambda_5$ allows larger maximum values of 
$M_h$.
For $\lambda_5=15$ one finds a horizontal line at $M_h\approx 670$ 
GeV,
showing that the upper bound has been increased for all values of
$\tan\beta$. In addition, large ($\ge 30$) values of $\tan\beta$
are allowed if $\lambda_5\ne 0$, in contrast to the case
of $\lambda_5=0$. For example, $\lambda_5=1$
comfortably permits values of $\tan\beta=60$.
However, as pointed out in \cite{HW} perturbative constraints on the
$\lambda_i$ also restrict the
allowed values of $\tan\beta$ in the THDM. Using the condition in
\cite{okada} which requires $|\lambda_i|\le 8\pi$, we found that
$\tan\beta\ge 30$ is strongly disfavoured.

We note that the relaxation of the strong correlation between $M_h$ and
$\tan\beta$ with $\lambda_5\ne 0$ would in principle allow the 
possibility
of distinguishing between the discrete symmetry conserving and violating 
potentials. If $h^0$ is discovered and the measured values of $M_h$
and $\tan\beta$ lie outside the rather constrained region for
$\lambda_5=0$, this would signify $\lambda_5\ne 0$ and thus a soft
breaking of the discrete symmetry.

\section*{5. Conclusions}
We have derived upper limits  on the masses of the Higgs bosons
both in the Minimal Standard Model (MSM) and also in
the general Two Higgs Doublet Model (THDM) by requiring that unitarity
is not violated in a variety of scattering processes.

In the MSM we have revived the unitarity constraints and shown
that the inclusion of all the coupled scattering channels
in the analysis can make the upper limit on the Higgs boson
stronger than in the non-coupled case.
The same analysis has been done for the THDM, where we first considered
the THDM scalar potential which is invariant under a discrete symmetry
transformation and improved previous studies by including the complete
set of scattering channels. Stronger constraints on the Higgs masses
were derived. Of particular interest is the $\tan\beta$ dependence
of the upper bound on $M_h$, with larger $\tan\beta$ requiring
a lighter $h^0$ e.g. $\tan\beta\ge 7$ implies
$M_h\le 100$ GeV. We then showed that the presence of the discrete 
symmetry breaking term parametrized by $\lambda_5$ may
significantly weaken the upper bounds on the masses.
In particular, the aforementioned correlation
between $\tan\beta$ and maximum $M_h$ is relaxed. It was suggested that
a measurement of $\tan\beta$ and $M_h$ may allow discrimination between
the two potentials.

\section*{Acknowledgements}
The work reported here concerning the THDM has been done in an enjoyable
collaboration with Andrew Akeroyd and ElMokhtar Naimi,
who are gratefully acknowledged.

\newpage
\renewcommand{\theequation}{A.\arabic{equation}}
\setcounter{equation}{0}
\subsection*{Appendix A.}
\subsection*{A.1 Quartic scalar interactions in the MSM}
The pure quartic scalar interactions in the MSM can be found from
the scalar potential $V=-\mu^2 \Phi^+ \Phi +\lambda (\Phi^+ \Phi)$
where $\Phi=\left( \begin{array}{c}
G^\pm \\ v + (\phi^0+i G^0 )/\sqrt{2}
\end{array}\right)$, $G^\pm$ and $G^0$ are
the goldstone boson and $\phi^0$ is the physical Higgs. Keeping only
the quartic terms in $V$ we get the following Feynman rules:
\begin{eqnarray}
& & G^+G^-G^+G^- =\frac{M_{\phi^0}^2}{v^2}  \quad , \quad
G^+G^-G^0 G^0 =\frac{3}{2}\frac{M_{\phi^0}^2}{v^2}  \quad , \quad
 G^+G^- \phi_0\phi_0 =\frac{M_{\phi^0}^2}{2v^2} \nonumber  \\
& & G^0G^0G^0G^0 =\frac{3}{2}\frac{M_{\phi^0}^2}{v^2}  \quad , \quad
\phi_0\phi_0 G^0 G^0 =\frac{M_{\phi^0}^2}{2v^2}  \quad , \quad
\phi_0\phi_0\phi_0\phi_0 = \frac{3}{2}\frac{M_{\phi^0}^2}{v^2}\nonumber 
\end{eqnarray}
where $v^2$ is fixed by the electroweak scale by
$v^2=(2\sqrt{2} G_F)^{-1}$, $M_{\phi^0}^2=4 v^2 \lambda$ and
$v^2=\frac{\mu^2}{4\lambda}$.
\subsection*{A.2 Quartic scalar interactions in the THDM}
We start from the potential \ref{higgspot}
in which the Higgs doublets
are expressed in terms of the non-physical fields:
$$\Phi_i=\left(
\begin{array}{c}
w_i^+\\
v_i+\frac{1}{\sqrt{2}}(h_i + iz_i)
\end{array}
\right)$$

Expanding this potential and keeping only the quartic terms
leads to the following feynman rules:

\begin{eqnarray}
\begin{array}{cc}
  w_1^+ w_1^- w_1^+ w_1^- = 4 (\lambda_1 + \lambda_3)\quad , & \quad
w_1^+ w_1^+ w_2^- w_2^- = (\lambda_5- \lambda_6)\nonumber  \\ [0.12cm]
  w_1^- w_1^+ w_2^- w_2^+ =
2\lambda_3+\frac{1}{2}( \lambda_5+\lambda_6) \quad , &  \quad
w_1^- w_1^- w_2^+ w_2^+ = (\lambda_5 - \lambda_6) \nonumber  \\ [0.12cm]
 w_2^+ w_2^- w_2^+ w_2^- = 4 (\lambda_2 + \lambda_3) \quad , & \quad
 h_1 h_1 h_1 h_1 = 6 (\lambda_1 +\lambda_3) \nonumber \\ [0.12cm]
h_2 h_2 h_2 h_2 = 6 (\lambda_2 + \lambda_3) \quad , & \quad
 h_1 h_1 h_2 h_2 = 2\lambda_3+\lambda_5 \nonumber\\ [0.12cm]
h_1 h_1 w_1^- w_1^+ = 2 (\lambda_1 + \lambda_3) \quad , & \quad
 h_2 h_2 w_1^- w_1^+ = 2\lambda_3+ \lambda_4\nonumber \\ [0.12cm]
h_1 h_2 w_1^+ w_2^- =  (\lambda_5-\lambda_4)/2  \quad , & \quad
 h_2 h_2 w_2^- w_2^+ = 2(\lambda_2+\lambda_3)\nonumber \\ [0.12cm]
h_1 h_2 w_1^- w_2^+ =  (\lambda_5-\lambda_4)/2 \quad , & \quad
 h_1 h_1 w_2^- w_2^+ =  2\lambda_3 + \lambda_4 \nonumber \\ [0.12cm]
 z_2 h_1 w_1^+w_2^- =\frac{i}{2} (\lambda_6-\lambda_4) \quad , & 
\quad
z_2 z_2 w_1^+w_1^- = 2\lambda_3+\lambda_4 \nonumber 
\end{array}
\end{eqnarray}

\begin{eqnarray}
\begin{array}{cc}
 z_2 h_1 w_1^-w_2^+ =-\frac{i}{2} (\lambda_6-\lambda_4) \quad , &
\quad
z_2 z_2 w_2^+w_2^- = 2(\lambda_2+\lambda_3) \nonumber \\ [0.12cm]
 z_2 z_2 h_2 h_2 = 2 (\lambda_2+\lambda_3) \quad ,&  \quad
 z_2 z_2 z_2 z_2 = 6 (\lambda_2+\lambda_3)\nonumber \\ [0.12cm]
 z_1 z_1 z_1 z_1 = 6 (\lambda_1 + \lambda_3) \quad , & \quad
 z_2 z_2 z_1 z_1 = 2\lambda_3+\lambda_5 \nonumber \\ [0.12cm]
 z_1 z_1 h_2 h_2 =  2\lambda_3+\lambda_6 \quad , & \quad
 h_1 h_1 z_1 z_1 = 2(\lambda_1+\lambda_3) \nonumber \\ [0.12cm]
 h_1 h_1 z_2 z_2= 2\lambda_3 +\lambda_6\quad , & \quad
 z_1 h_2 w_1^+ w_2^- = - \frac{i}{2} (\lambda_6-\lambda_4) 
\nonumber \\ [0.12cm]
z_1 h_2 w_1^- w_2^+ =  \frac{i}{2} (\lambda_6-\lambda_4) \quad , & \quad
 z_1 z_1 w_1^+ w_1^- = 2 (\lambda_1+\lambda_3) \nonumber \\ [0.12cm]
z_1 z_1 w_2^- w_2^+ =2 \lambda_3 + \lambda_4 \quad , & \quad
 z_1 z_2 h_1 h_2 = (\lambda_5 - \lambda_6)/2 \nonumber\\ [0.12cm]
z_1 z_2 w_1^+ w_2^- = (\lambda_5 - \lambda_4)/2 \quad & \quad
 z_1 z_2 w_1^- w_2^+ = (\lambda_5 - \lambda_4)/2\nonumber
\end{array}
\end{eqnarray}

\end{document}